\documentclass[sigconf]{acmart}
\usepackage{booktabs} 
\usepackage{graphicx}
\usepackage{xcolor}
\usepackage{xspace}
\usepackage[normalem]{ulem}
\useunder{\uline}{\ul}{}
\usepackage{multirow}
\usepackage{lscape}
\usepackage{rotating}
\usepackage{supertabular}
\usepackage{longtable}
\usepackage{enumerate}
\newif\ifdraft
\ifdraft
  \newcommand{\jon}[1]{{\color{blue}\emph{Jon: #1}}\xspace}
  \newcommand{\arif}[1]{{\color{orange}\emph{Arif: #1}}\xspace}
  \newcommand{\davoud}[1]{{\color{magenta}\emph{Davoud: #1}}\xspace}
  \newcommand{\waqar}[1]{{\color{red}\emph{Waqar: #1}}\xspace}
  \newcommand{\harsha}[1]{{\color{brown}\emph{Harsha: #1}}\xspace}
  \newcommand{\rifat}[1]{{\color{violet}\emph{Rifat: #1}}\xspace}
\else
  \usepackage[disable]{todonotes}
  \newcommand{\jon}[1]{}
  \newcommand{\arif}[1]{}
  \newcommand{\davoud}[1]{}
  \newcommand{\waqar}[1]{}
  \newcommand{\harsha}[1]{}
  \newcommand{\rifat}[1]{}
\fi
\usepackage{hyperref}
\acmConference[OVIS]{}{}{Monash University, Australia}

\renewcommand\footnotetextcopyrightpermission[1]{} 

\usepackage{enumitem}
\begin{document}


\title[A Study on the Prevalence of Human Values]{A Study on the Prevalence of Human Values \\in Software Engineering Publications, 2015 -- 2018}




\author{Harsha Perera}
\affiliation{%
  \institution{Monash University}
  \streetaddress{}
  \city{Clayton}
  \state{Australia}
  \postcode{3800}
}
\email{harsha.perera@monash.edu}

\author{Arif Nurwidyantoro}
\orcid{}
\affiliation{%
  \institution{Monash University}
  \streetaddress{}
  \city{Clayton}
  \state{Australia}
  \postcode{3800}
}
\email{arif.nurwidyantoro@monash.edu}

\author{Waqar Hussain}
\orcid{}
\affiliation{%
  \institution{Monash University}
  \streetaddress{}
  \city{Clayton}
  \state{Australia}
  \postcode{3800}
}
\email{waqar.hussain@monash.edu }

\author{Davoud Mougouei}
\orcid{}
\affiliation{%
  \institution{Monash University}
  \streetaddress{}
  \city{Clayton}
  \state{Australia}
  \postcode{3800}
}
\email{davoud.mougouei@monash.edu }

\author{Jon Whittle}
\authornote{Prof. Jon Whittle is the leader of OVIS (Operationalizing Human Values)  Lab in the Faculty of Information Technology at Monash University.}
\orcid{}
\affiliation{%
  \institution{Monash University}
  \streetaddress{}
  \city{Clayton}
  \state{Australia}
  \postcode{3800}
}
\email{jon.whittle@monash.edu}

\author{Rifat Ara Shams}
\orcid{}
\affiliation{%
  \institution{Monash University}
  \streetaddress{}
  \city{Clayton}
  \state{Australia}
  \postcode{3800}
}
\email{rifat.shams@monash.edu} 

\author{Gillian Oliver}
\orcid{}
\affiliation{%
  \institution{Monash University}
  \streetaddress{}
  \city{Clayton}
  \state{Australia}
  \postcode{3800}
}
\email{gillian.oliver@monash.edu}

\renewcommand{\shortauthors}{H. Perera et al.}

%
%


\begin{abstract}
Failure to account for human values in software (e.g., equality and fairness)\waqar{don't need a comma after fairness} can result in user dissatisfaction and negative socio-economic impact. Engineering these values in software,\waqar{one comma after however should be fine} however, requires technical and methodological support throughout the development life cycle. This paper investigates to what extent software engineering (SE) research has considered human values. We investigate the prevalence of human values in recent (2015 -- 2018) publications at some of the top-tier SE conferences and journals. We classify SE publications, based on their relevance to different values, against a widely used value structure adopted from social sciences. Our results show that: (a) only a small proportion of the publications directly consider values, classified as \textit{relevant publications}; (b) for the majority of the values, very few or no relevant publications were found; and (c) The prevalence of the relevant publications was higher in SE conferences compared to SE journals. This paper shares these and other insights that motivate research on human values in software engineering.

\waqar{for consistency in the use of tenses we can either use
"we have investigated" , we have classified"
, our results have shown 
or 
"We investigated", "we classified", "our results showed"
my favourite is 
We investigate, we classify, and our results show"}
\end{abstract}
\maketitle
\section{Introduction}
\label{sec:introduction}

Ignoring human values while engineering software may \waqar{you are using 'may result' here (which shows a possibility. Use consistently both in the abstract and intro) here in the intro, where as you used 'results' (a certainty) in the abstract}result in violating those values~\cite{Mougouei2018,ferrario2016values} and subsequent dissatisfaction of users. This may lead to negative socio-economic impacts such as financial loss and reputational damage\waqar{not sure which organizations we mean here}. A recent example, which made news headlines, is the price gouging on airline tickets during Hurricane Irma~\cite{sablich_2017}. After a mandatory evacuation order, the cost of airline tickets rose six fold, due to supply and demand pricing systems, thus disadvantaging evacuees. Arguably, this occurred because of insufficient consideration of valuing compassion for those suffering in a natural disaster.\waqar{if I recall Jon commented regarding this example that it needs to be somehow linked with the value of compassion, i think it still needs that kind of link and motivation} A second example is software used by Amazon to determine free shipping by zip code, which turned out to discriminate against minority neighbourhoods~\cite{gralla_2016}. Racial bias in automatic prediction of re-offenders at parole boards in the US Justice system~\cite{angwin_larson_kirchner_mattu_2016} is another example where software violates human values\waqar{values violation}. Indeed, the negative impacts of ignoring values can go as far as risking human life: the tragic suicide of the British teenager Molly Russell~\cite{molly.2019} has been partially attributed to  Instagram's personalisation algorithms, which flooded Molly's feed with self harm images; following public outrage\waqar{accusations}, Instagram has now banned such images.

As awareness about human aspects\waqar{aspects?} of software grows, the public is increasingly demanding software that accounts for their values. See, for example, those accusing Facebook of taking advantage of users' data to influence the US elections\waqar{rephrase: not clear} \cite{smith_2018}. Public demand has also motivated software vendors to take preemptive measures to avoid violating human values. Google, for instance, has pledged not to use its AI tools for surveillance conflicting with human rights~\cite{dave_2018}. 
 
Though such initiatives are promising, we claim that software engineering research and practice currently pays insufficient attention to the majority of human values. This may be due to the lack of adequate methodological and technical support for engineering values in software~\cite{Mougouei2018}. To provide evidence for this claim, as part of our broader approach to studying human values, we have investigated software engineering (SE) research papers\waqar{the justification, why study SE research to answer problems in SE practice, is missing} to measure how much attention the SE field has given to values. In particular, we have classified software engineering publications in some of the top-tier SE venues (ICSE, FSE, TSE, and TOSEM), from 2015 to 2018, based on their relevance to different values. A paper was classified as \textit{directly relevant} to a particular value if its main research contribution addressed how to define, refine, measure, deliver or validate this value in software\waqar{what do you mean by directly considered a value}. A widely adopted value structure (Figure~\ref{fig:values}), based on Schwartz's theory of human values~\cite{Schwartz2012,schwartz1992universals}, was used as our classification scheme\waqar{do we need to say 'primary'}. \waqar{we don't need 'in the effort' we can start from To understand }Using this classification approach, we investigated the prevalence of human values in SE research, with three key research questions:


\begin{itemize}
    \item [\textbf{(RQ1)}] To what extent are SE publications relevant to values? 
    \item [\textbf{(RQ2)}] Which values are commonly considered in SE publications?
    \item [\textbf{(RQ3)}] How are the relevant publications distributed across venues? 
\end{itemize}

 
The results of our study showed that: (a) only $16\%$ of publications were directly relevant to human values, referred to, henceforth, as \textit{relevant publications}; (b) for $60\%$ of human values, there were no relevant publications; (c) on average, $2$ relevant papers were found per value\waqar{which each value?}, while for $79\%$ of values, the number of relevant publications was $\leq 2$; and (d) $88\%$ of relevant papers\waqar{from 2015 to 2018, or during the selected period i.e. 2015-2018} were published in SE conferences rather than journals. 

\section{Background}
\label{sec:background}

Cheng and Fleischmann summarize seven different definitions of human values as ``guiding principles of what people consider important in life''~\cite{cheng2010developing}. Human values with an ethical and moral import such as \textit{Equality}, \textit{Privacy} and \textit{Fairness} have been studied in technology design and human-computer interaction for more than two decades \cite{friedman1996value,flanagan2005values,friedman2007human}. Meanwhile, the rapid popularization of artificial intelligence (AI) and its potential negative impact on society have raised the awareness of human values in AI research~\cite{riedl2016using, etzioni2017incorporating,cath2018artificial}. Consequently, human values are getting renewed research focus. 
 
 There has been some recent (but isolated) research  in software engineering such as values-based requirements engineering \cite{thew2018value}, values-first SE \cite{ferrario2016values} and values-sensitive software development \cite{aldewereld2015design}.
However, there has been no previous work that measures to what extent human values have been considered in SE research. Motivated by this research gap, we follow a classification approach, similar to that used in previous SE research to map topic trends \cite{shaw2003writing,systa2012inbreeding,montesi2008software}, but with a different purpose, to measure values relevance. There are no current classification schemes for human values in SE. Therefore, we take inspiration from the social sciences.

Social scientists have been searching for the most useful way to conceptualize basic human values since the 1950s \cite{schwartz2007basic}. In 1973, Rockeach captured 36 human values and organized them into 2 categories \cite{rokeach1973nature}. In 1992, Schwartz introduced his theory of basic human values (henceforth referred to as Schwartz's Values Structure (SVS)) which recognized 58 human values categorized into 10 value categories \cite{schwartz1992universals, schwartz2005basic}. While these two value structures remain the most well recognized ways of representing values, there are at least ten other value classifications \cite{cheng2010developing}. In this paper, we use SVS, which is  the most cited and most widely applied classification not only in the social sciences but also in other disciplines \cite{thew2018value, Ferrario2014}.\harsha{need more references here}



\begin{figure*}[!htbp]
	\centering
	\centerline{\includegraphics[scale=0.75,angle=0]{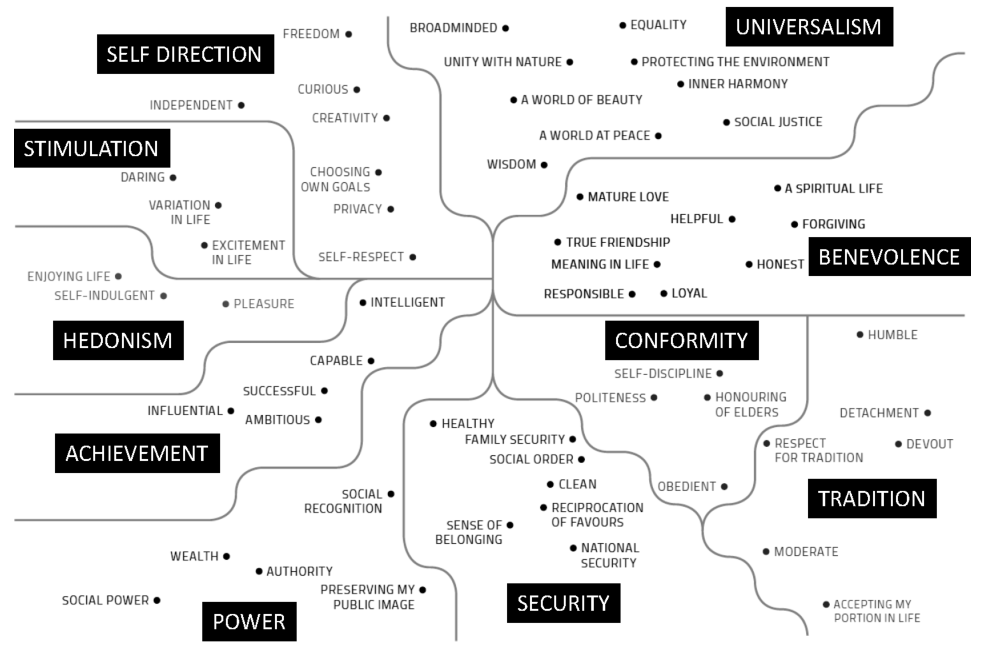}}
	\captionsetup{margin=3cm}
	\caption{Schwartz Values Structure~\cite{schwartz2006valeurs,schwartz2004evaluating} (adopted from \cite{holmes_blackmore_hawkins_wakeford_2011}). Words in black boxes are values categories, each subdivided into values.}
	\label{fig:values}
\end{figure*}

In SVS, Schwartz introduced 10 motivationally-distinct value categories recognized across more than 30 cultures \cite{schwartz1992universals}. Each value category has underlying distinct motivational goals (see Table \ref{tab:valuecategories_defOnly}) which relate to three fundamental needs of human existence \cite{schwartz1992universals}. 
 Schwartz subdivided each value cateogory into a set of closely related values \cite{schwartz1994there, schwartz1992universals}.
These 10 value categories and 58 values are arranged in a circular motivational structure as shown in Figure \ref{fig:values}. Value categories located close to each other are complementary whereas values further apart tend to be in tension with each other. Section \ref{sec:methodology} discusses how we applied SVS in our classification study.

\begin{table}
\caption{Value categories and descriptions \cite{Schwartz2012} \harsha{\davoud{the reader is not a social scientist, we are writing for software engineers => use language that they understand; \harsha{this is adopted directly from the source, should I rephrase, I don't think we should as these were defined in their publications}}}}
\label{tab:valuecategories_defOnly}
\centering
\small
\begin{tabular}{lp{5.3cm}}
\toprule
\textbf{Value Category} & \textbf{Description (motivational goals)} \\ \midrule
Self-direction & Independent thought and action--choosing, creating, exploring \\ \hline
Stimulation & Excitement, novelty, and challenge in life\\ \hline
Hedonism & Pleasure or sensuous gratification for oneself\\ \hline
Achievement & Personal success through demonstrating competence according to social standards\\ \hline
Power & Social status and prestige, control or dominance over people and resources\\ \hline
Security & Safety, harmony, and stability of society, of relationships, and of self\\ \hline
Conformity & Restraint of actions, inclinations, and impulses likely to upset or harm others and violate social expectations or norms\\ \hline
Tradition & Respect, commitment, and acceptance of the customs and ideas that one's culture or religion provides\\ \hline
Benevolence & Preserving and enhancing the welfare of those with whom one is in frequent personal contact\\ \hline
Universalism & Understanding, appreciation, tolerance, and protection for the welfare of all people and for nature \\ \jon{\hline
Holistic View & Human values considered holistically without focusing on predetermined values\\ }
\bottomrule
\end{tabular}
\end{table}

\section{Methodology}
\label{sec:methodology}

    To investigate the prevalence of human values in SE research, we manually classified publications from top-tier SE conferences and journals based on their relevance to different values. We followed a methodology similar to that of prior classification work in SE \cite{shaw2003writing,systa2012inbreeding,montesi2008software}, which mapped trends of SE research over time in terms of topic and type of study. As with prior studies, ours was based on manual classification of paper abstracts by multiple raters. Classification based on abstracts, rather than reading the full paper, is sub-optimal but strikes a balance between accuracy and time needed for the study. All papers had multiple raters and inter-rater agreement was measured using Fleiss' Kappa \cite{landis1977measurement}. We chose to classify papers from the last four years of conferences and journals generally considered to be the top SE venues, namely, the International Conference on Software Engineering (ICSE), the ACM Joint European Software Engineering Conference and Symposium on the Foundations of Software Engineering (ESEC/FSE), the IEEE Transactions on Software Engineering (TSE), and the ACM Transactions on Software Engineering and Methodology (TOSEM). 
    
    When conducting such a study, there are a number of key experimental design decisions that need to be taken, including: (i) how to define relevance to human values, given the imperfect and high-level nature of values definitions in the literature; (ii) how many raters to assign to each paper, and (iii) how to resolve disagreements between raters. To make choices about these design decisions, we first carried out a pilot study before carrying out the main study. Both the pilot and main study assumed SVS as the classification scheme. In total, we employed 7 raters (5 Male, 2 Female) with varying levels of  experience in SE research, ranging from PhD students to  Professors, and including one rater from outside the software engineering field. Note that this is a  relatively high number of raters compared to similar studies~\cite{Bertolino2018, vessey2002research}.





\subsection{Pilot Study}
\label{subsec:Pilot-Phase}
The pilot study had three steps: (i) Paper selection and allocation of papers to raters, (ii) Paper classification, and (iii) Calibration of classification decisions made by different raters. 
The aim of the pilot study was not to measure relevance of papers to values; rather, we had the following objectives:
\begin{itemize}
  \item To test the appropriateness of SVS as the classification scheme for SE publications
  \item To develop a common understanding regarding the meaning of human values in SE contexts
   \item To collect insights from raters to feed into the experimental design of the main study
\end{itemize}

\textit{(i) Paper selection and allocation of papers to raters.} We randomly selected 49 papers from ICSE 2018 as our pilot study dataset. These were equally allocated among the seven raters, with three raters per paper. Common practice is to assign two raters per paper \cite{Bertolino2018, vessey2002research}; three were assigned in the pilot to get a better understanding of how to map papers to values. ICSE was chosen as it has the broadest coverage of SE research \cite{Bertolino2018}. We chose the most recent ICSE proceedings -- 2018 at time of writing. 

\textit{(ii) Paper classification.} Raters classified papers, independently, based on their title, abstract and keywords which is an approach used in similar classification studies in SE \cite{shaw2003writing, glass2002research, Bertolino2018}. Raters were instructed to decide if a paper was ``relevant'' or ``not relevant'' to human values: relevance was deliberately left ill-defined as one of the objectives of the pilot was to influence the definition of this term in the main study. For relevant papers, raters were asked to classify the papers into one value category, and then into one value within the category. Raters were not mandated to follow the hierarchical structure of SVS: that is, they could classify a paper into value X and value category Y even if X did not belong to category Y. This was to give us a way to assess, from a software engineering perspective, the appropriateness of the hierarchy in SVS.

\textit{(iii) Calibration.} After classification, all seven raters met to discuss the classification decisions. The main objective was to calibrate decisions and use this to refine the definition of values relevance. The intention was \textit{not} to decide which rater picked the correct classification.

Following the pilot study, we made a number of observations which were fed into experimental design of the main study.

\begin{itemize}[leftmargin=0.3cm]

\item \textit{Observation 1:} Raters found that almost every paper could be classified into a small number of values such as \textit{Helpfulness}, \textit{Wisdom} or \textit{Influence} because, in general, every piece of research tries to advance knowledge. Thus, an indirect argument could almost always be made why a paper is relevant to helpfulness (e.g., a paper on testing is helpful to testers), wisdom (any paper advances knowledge, thus leading to greater wisdom), or influence (e.g., a paper on an improved software process influences how software is developed). This observation illustrated the difficulty of working with vaguely defined concepts such as values, but also the importance of a better definition of relevance.

\item \textit{Decision 1:} It is beyond the scope of this paper to fully and formally define all the values; hence, it was decided in the main study to use inter-rater agreement as evidence that a value was sufficiently understood in the context of a particular paper to provide confidence in the results. The definition of relevance was, however, refined for the main study. Raters were instructed not to make indirect arguments why a paper might be relevant to a value. Instead, in the main study, classification was based on ``direct relevance'' -- a paper is defined as directly relevant to a value if its main research contribution is to define, refine, measure, deliver or validate a particular value in software development. All other papers are classified as not relevant. Thus, a paper should only be classified as directly relevant to helpfulness if the research provides software tools or techniques to encourage people to be helpful towards each other. 

\item \textit{Observation 2:} Raters observed that some papers addressed values as a general concept rather than considering any specific value. An example would be a paper that presents a methodology for refining values into a software architecture. These papers should not be classified into any particular value category or value.

\item\textit{Decision 2:} To facilitate classification of such papers, we introduced a new value category in the main study, named \textit{Holistic view}. A paper classified under Holistic View relates to values generally without focusing on any specific value (Table \ref{tab:example}).

\item \textit{Observation 3:} Raters found that some papers should be classified under more than one value. 

\item \textit{Decision 3:} To accommodate such papers in the main study, raters were allowed to select up to three values.\jon{why 3} This decision is different from similar  studies in SE where raters were obliged to pick just one category \cite{Bertolino2018}. 

\item \textit{Observation 4:} Not surprisingly, as SVS was not developed specifically for SE, there were cases where  SVS  was not a perfect fit. We will return to this point in Section \ref{sec:conclusion} but a key point for the main study is that some raters chose a value X and value category Y even if X does not belong to Y according to SVS. A common example was the value \textit{Privacy}, which from a SE perspective is clearly aligned with the category \textit{Security}, and yet appears in \textit{Self-direction} according to SVS (see Figure \ref{fig:values}).

\item \textit{Decision 4:} The main study maintained the decision to allow selection of values and value categories independent of the Schwartz hierarchical structure. In Section \ref{sec:result}, we present data to show the effect this had on the results.


\item \textit{Observation 5:} The pilot study gave us an opportunity to measure how long it took raters to rate papers. We found that, on average, each rater spent four minutes per abstract. Given the number of papers in the main study (1350 -- see Table \ref{tab:papercount}), assigning three raters per paper would be infeasible.

\item \textit{Decision 5:} Out of necessity, we reduced the number of raters in the main study to two. This is consistent with the number of raters in similar studies \cite{Bertolino2018, vessey2002research, glass2002research}.
\end{itemize}


\subsection{Main Study}
\label{subsec:classification-process}

Similar to the pilot study, the main study also had three phases: (i) Paper selection and allocation of papers to raters, (ii) Paper classification and (iii) Disagreement resolution. The final stage was different to the pilot study because rather than calibrating ratings to inform experimental design, some raters met to try and reach a consensus.

\textit{(i) Paper selection and allocation of papers to raters.} For the main study, we selected papers from ICSE, FSE, TSE and TOSEM over the last four years. These are the same venues used in similar paper classification studies \cite{Bertolino2018, glass2002research}. We selected all papers in TSE and TOSEM. For FSE, we used all papers from the main track, and for ICSE, we used all papers from the main track, from the Software Engineering in Practice (SEIP) track, and from the Software Engineering in Society (SEIS) track. The motivation for selecting tracks was to choose tracks which publish full research papers, not shorter papers. In total, there were 1350 papers published in the chosen venues over the years 2015--2018, at time of writing. This is a high sample size compared to similar studies (e.g., 976 in Bertolino et al. \cite{Bertolino2018} and 369 in Glass \cite{glass2002research}). Table \ref{tab:papercount} shows the distribution of selected papers by venue, track and year.
\begin{table}
\caption{Classified publications by venue/track and year}
\label{tab:papercount}
\resizebox{0.46\textwidth}{!}{  
\small
\begin{tabular}{l|l|l|l|l|l}
\hline
\multicolumn{1}{l|}{\textbf{Venue \& Track}} & \multicolumn{1}{l|}{\textbf{2015}} & \multicolumn{1}{l|}{\textbf{2016}} & \multicolumn{1}{l|}{\textbf{2017}} & \multicolumn{1}{l|}{\textbf{2018}} & \multicolumn{1}{l}{\textbf{Total}} \\ \hline
ICSE--Main Track                     & 83                        & 101                       & 68                        & 153                       & 405                        \\ \hline
ICSE--SEIP                           & 25                        & 28                        & 30                        & 35                        & 118                        \\ \hline
ICSE--SEIS                           & 9                         & 7                         & 9                         & 11                        & 36                         \\ \hline
ESEC/FSE--Main Track                 & 123                       & 143                       & 124                       & 122                       & 512                        \\ \hline
TSE                                  & 62                        & 61                        & 61                        & 31                        & 215                        \\ \hline
TOSEM                                & 22                        & 16                        & 12                        & 14                        & 64                         \\ \hline
Total                                & 324                       & 356                       & 304                       & 366                       & 1350                       \\ \hline
\end{tabular}
}
\end{table}
  The papers were randomly allocated among the seven raters, two raters per paper. Each rater received around 400 papers to classify.   We manually extracted  links for each of the 1350 papers from digital databases, and provided a spreadsheet with these links and values and value categories for raters to select from. 

\textit{(ii) Paper classification.} 
Similar to the pilot study, raters were asked to classify papers on the basis of their title, abstract and keywords. However, the main study used a different definition of relevance, as suggested by the pilot study. Raters were asked to classify papers as directly relevant or not directly relevant, where the definition of direct relevance is as given in Section \ref{subsec:Pilot-Phase}. Papers found directly relevant to values were further classified into a category and then to a specific value(s). Throughout the process, raters complied with the decisions made during the \textit{calibration} step in the pilot study.

\textit{(iii) Disagreement resolution.} Given the subjective nature of the classification, raters sometimes disagreed. This could arise at three levels:
(a) relevance level, where raters disagreed on whether a paper was directly relevant or not; (b) value category level, where raters disagreed on the choice of value category; and (c) value level, where raters disagreed on the choice of value.



To attempt to resolve these disagreements, raters met to discuss their views about why the paper in question was classified in a certain way.  If the raters could not come to an agreement, a third rater was introduced as an arbiter. The arbiter facilitated a second round of discussion, sharing his or her own views, to facilitate a consensus. However, if the disagreement persisted, the arbiter did not force a decision.




Aligned with previous studies \cite{Bertolino2018}, we calculated inter-rater agreement using Fleiss' Kappa, once attempts at resolving disagreements had taken place. The results of the Kappa measure are interpreted according to the agreement strengths introduced by Landis and Koch \cite{landis1977measurement}. We achieved \textit{almost perfect} agreements on relevance level and category level with Kappa values equal to 0.92 and 0.87, respectively. The agreement of value level was found as \textit{substantial} with Kappa value equal to 0.79.  The results from the main study are further discussed in Section \ref{sec:result}.

\begin{table*}[t]
\centering
\caption{Examples of paper classification at different levels (\textit{relevance}, \textit{value category}, and \textit{value}).\harsha{shall I add an example on privacy here for my reference}\davoud{sure, if we have enough space}}
\label{tab:example}
\small
\begin{tabular}{p{2cm}p{15cm}}
\toprule\multicolumn{1}{c}{\textbf{Classification}} & \textbf{Example}                                                                                                                                                              \\ \midrule
Not Relevant                       & CafeOBJ is a language for writing formal specifications for a wide variety of software and hardware systems and for verifying their properties ... we have extended CafeInMaude, a CafeOBJ interpreter implemented in Maude, with the CafeInMaude Proof Assistant (CiMPA) and the CafeInMaude Proof Generator (CiMPG) ... \cite{riesco2018prove}                                                                                        \\ \hline
Privacy & Network traffic data contains a wealth of information for use in security analysis and application development. Unfortunately, it also usually contains confidential or otherwise sensitive information, ... We present Privacy-Enhanced Filtering (PEF), a model-driven prototype framework that relies on declarative descriptions of protocols and a set of filter rules ... \cite{dijk2017model}     \\ \hline
Helpful                                     & ... However, newcomers face many barriers when making their first contribution to an OSS project, leading in many cases to dropouts. Therefore, a major challenge for OSS projects is to provide ways to support newcomers during their first contribution. In this paper, we propose and evaluate FLOSScoach, a portal created to support newcomers to OSS projects. ... \cite{Steinmacher:2016:OOS:2884781.2884806}                       \\ \hline
Protecting the Environment                  & ... The battery power limitation of mobile devices has pushed developers and researchers to search for methods to improve the energy efficiency of mobile apps. We propose a multiobjective refactoring approach to automatically improve the architecture of mobile apps, while controlling for energy efficiency... \cite{morales2018earmo}                                                                         \\ \hline
Holistic View                               & ... The aim of this paper is to give more visibility to the interrelationship between values and SE choices. To this end, we first introduce the concept of Values-First SE and reflect on its implications for software development. Our contribution to SE is embedding the principles of values research in the SE decision making process and extracting lessons learned from practice. ... \cite{ferrario2016values} \\ \bottomrule
\end{tabular}
\end{table*}

\vspace{1cm}
\section{Results}
\label{sec:result}

This section presents the results of the main study described in Section \ref{subsec:classification-process}. As a reminder, we  investigate the following research questions:

\begin{itemize}
    \item [\textbf{(RQ1)}] To what extent are SE publications relevant to values? 
    \item [\textbf{(RQ2)}] Which values are commonly considered in SE publications?
    \item [\textbf{(RQ3)}] How are the relevant publications distributed across venues? 
\end{itemize}

\subsection{The Prevalence of Values in SE Publications}
\label{subsec:values-prevalence}

To answer \textbf{(RQ1)} and \textbf{(RQ2)}, in this section, we present the results achieved from Section~\ref{subsec:classification-process}  and discuss our findings on the prevalence of human values in SE Publications. 

\subsubsection{Answering \textbf{(RQ1)}} 

\begin{figure}[htb]
    \centering
    \includegraphics[width=0.48\textwidth]{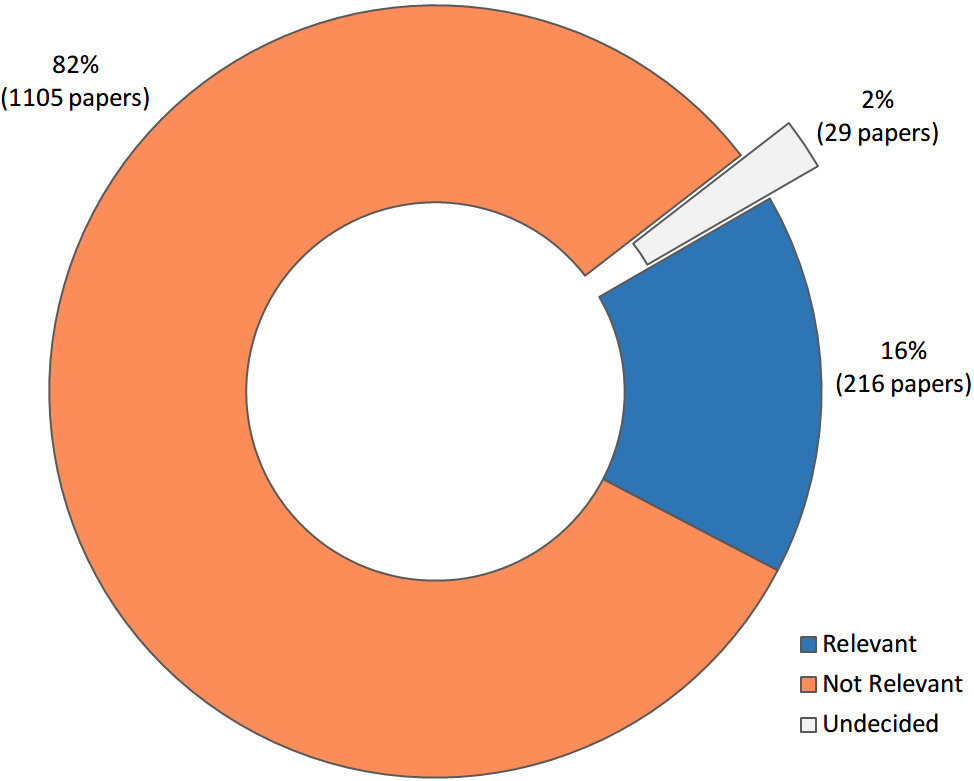}
    \caption{Relevance of SE publications to human values}
    \label{fig:relevant}
\end{figure}

\begin{figure}[htb]
    \centering
    \includegraphics[width=0.45\textwidth]{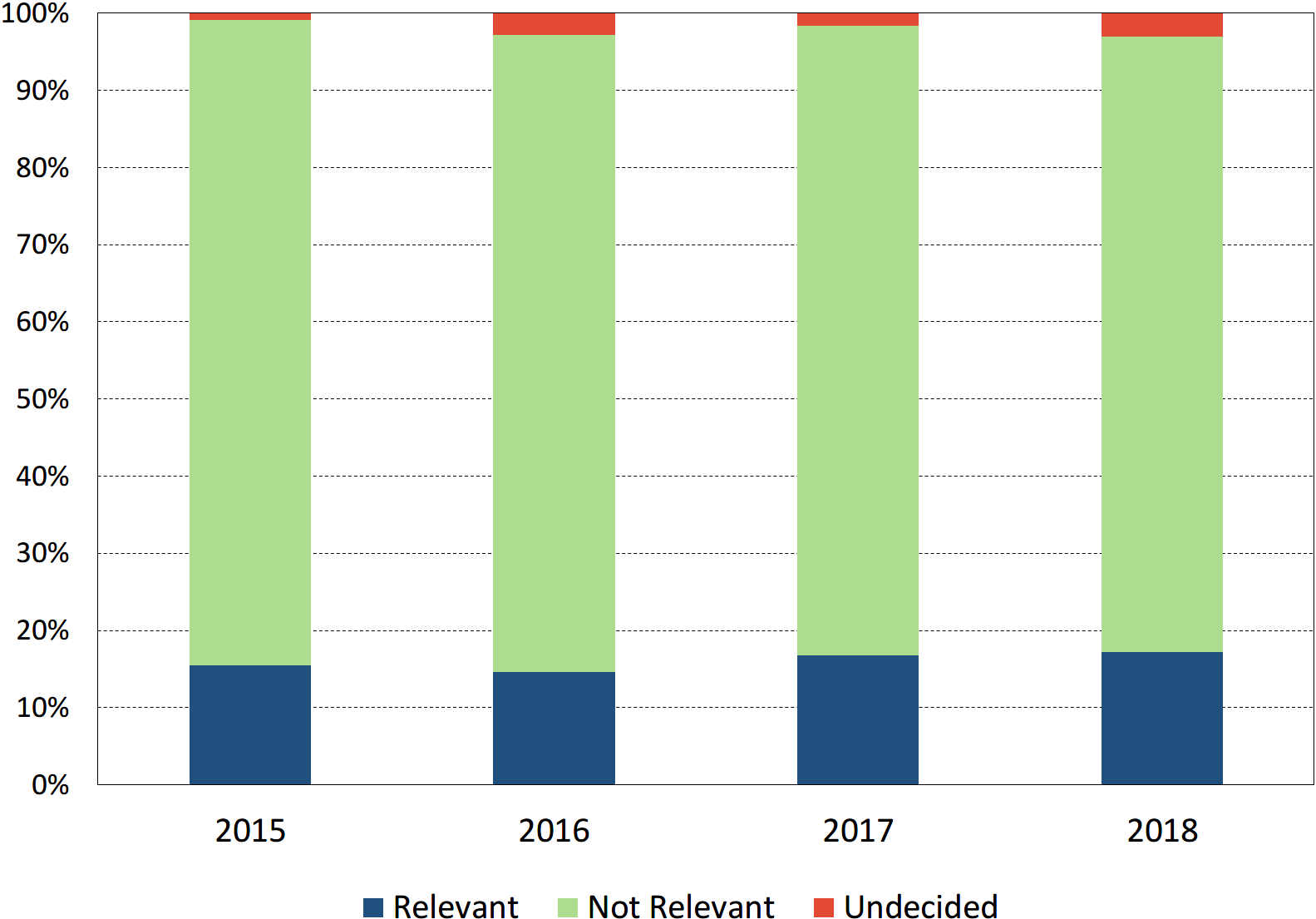}
    \caption{Relevant publications per year.}
    \label{fig:supporting-ratio-year}
\end{figure}

Figure~\ref{fig:relevant} demonstrates the prevalence of human values in classified publications. We observed (Figure~\ref{fig:relevant}) that the majority of the publications (82\%) were classified as \textit{Not Relevant} to values, which constitutes 1105 out of 1350 papers. For those publications that did not directly relate to values, an example is given in Table~\ref{tab:example}. On the other hand, 16\% of the publications (216 papers) were found to be directly relevant to values. The remaining 2\% of publications (29 papers) were classified as undecided, because the two raters could not agree on a classification. To investigate if there were any trends in the prevalence of  values in SE venues over time, we compared the percentages of the relevant publications from 2015 to 2018 (Figure~\ref{fig:supporting-ratio-year}): no significant trends were observed. 

It is worth mentioning that even though the raters agreed that 216 papers (16\% of the classified papers) were relevant to values, disagreements still remained at the value category level and value level (Section~\ref{subsec:classification-process}): out of 216 papers, agreements were reached for 195 papers at value category level and at the value level, agreements were reached for 115 papers. 

\subsubsection{Answering \textbf{(RQ2)}}

\textit{Which values are commonly considered?}

\begin{figure*}[htbp]
    \centering
    \hspace{2cm}\includegraphics[width=0.75\textwidth]{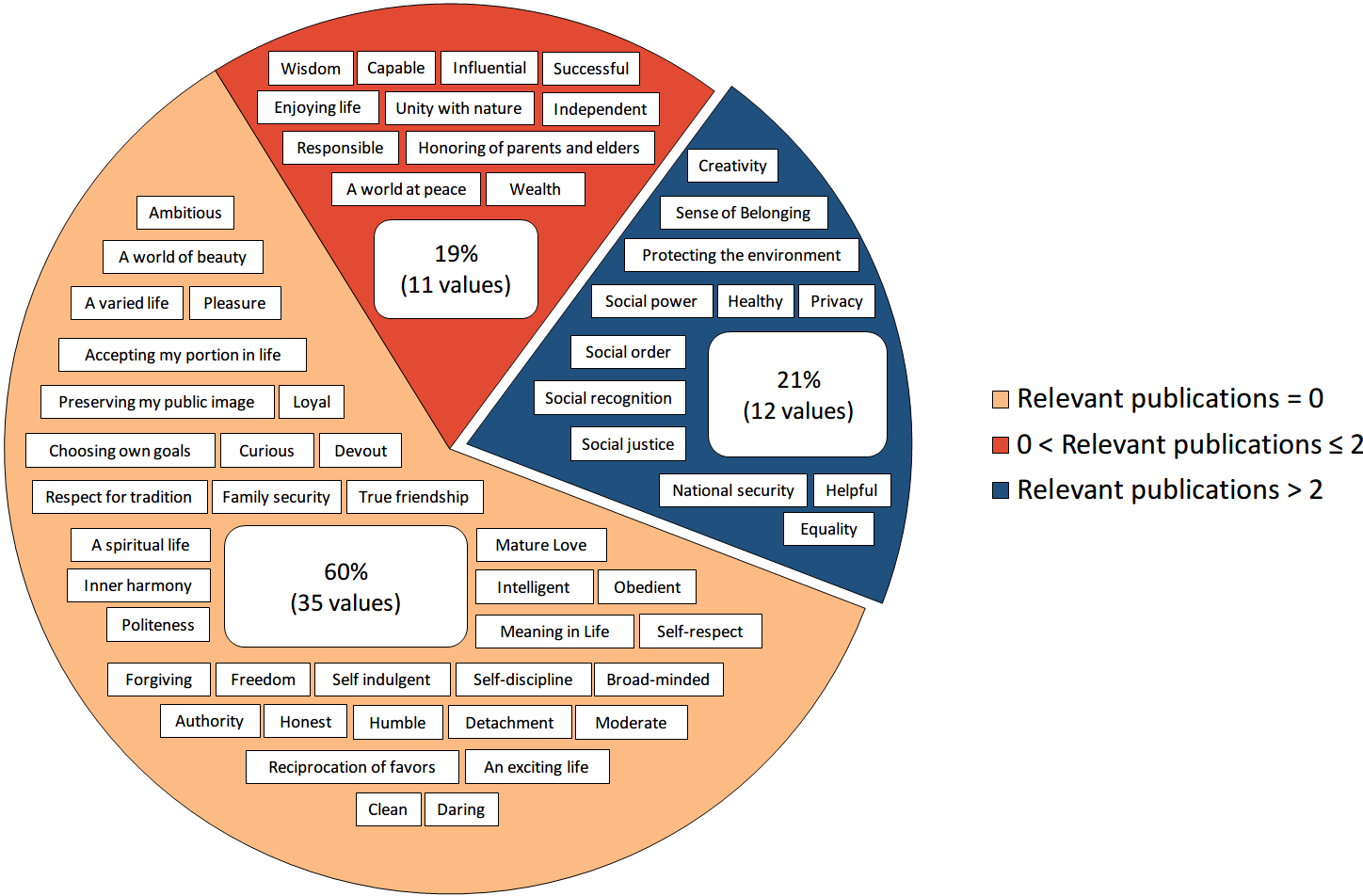} 
    \captionsetup{margin=15ex}
    \caption{The level of attention given to 58 values in the Schwartz Value Structure. Publications were classified as relevant if their main research contribution directly considered values.}
    \label{fig:specific-values}
\end{figure*}

\begin{figure}[htb]
    \centering
    \includegraphics[width=0.5\textwidth]{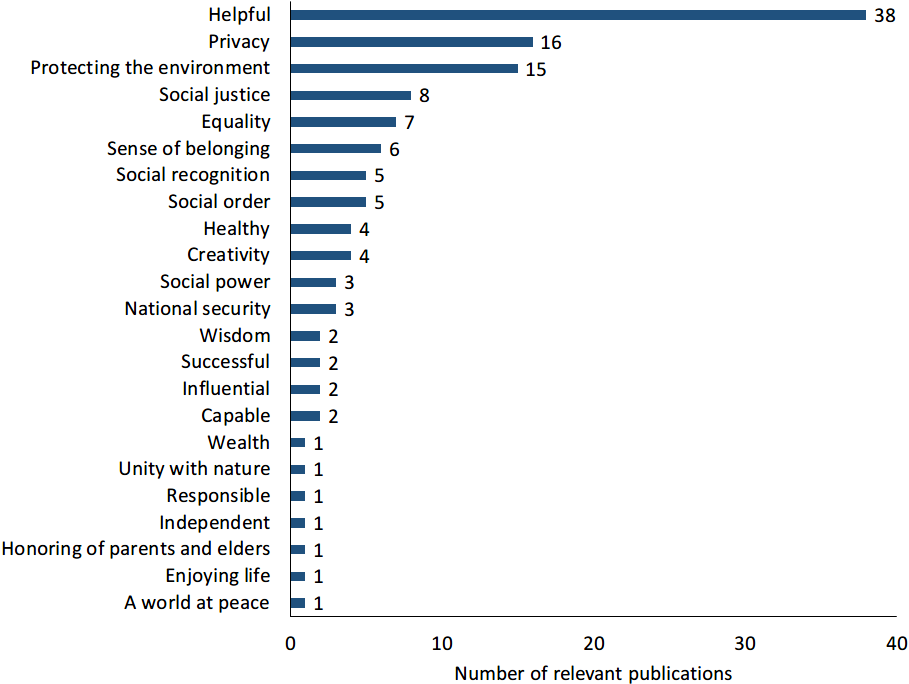}
    \caption{The number of relevant publications per value}    
    \label{fig:specific-values-occurrences}
\end{figure}

Our results showed that for each of the 58 values in Figure~\ref{fig:values} -- on average -- $2$ relevant publications were found\davoud{box plot ?}. As shown in Figure~\ref{fig:specific-values-occurrences}, however, the frequency of the relevant publications varied significantly for different values. Figure~\ref{fig:specific-values} shows the level of attention given to the 58 human values in SVS. 

It can be seen that for the majority of the values ($79\%$), the number of the relevant publications was $\leq 2$  while for 60\% (35 out of 58) of the values, no relevant publications were found (Figure~\ref{fig:specific-values}). Also, for some values, e.g., \textit{Enjoying life} and \textit{Honoring of parents and elders}, only one relevant publication was found across all of the studied venues from 2015 -- 2018 (Figure~\ref{fig:specific-values-occurrences}). It can also be seen in Figure~\ref{fig:specific-values} that only for 21\% (12 out of 58) of the values, e.g. \textit{Helpful} and \textit{Privacy}, the number of the relevant publications were above average ($> 2$).

While being cautious with generalizing, these findings are highly suggestive of negligible or limited attention paid by the SE research community to the majority of human values. Although finding the exact cause requires broader studies,  it may not be difficult to attribute ignoring some of the values in SE publications to the lack of practical definitions for those values~\cite{Mougouei2018}; this is particularly clear for values such as \emph{Forgiving} and \emph{Mature love}, that need to be further clarified in a SE context before they can be used by SE researchers and practitioners. 

In the attempt to understand which values are most commonly considered in SE research, we found (Figure~\ref{fig:specific-values-occurrences}) that the number of publications relevant to \emph{Helpful}, \emph{Privacy}, and \emph{Protecting the environment}, were the highest among all 58 values in SVS (Figure~\ref{fig:values}). Examples of such publications are given in Table~\ref{tab:example}. With 38 relevant papers, the value \textit{Helpful} was the most frequently considered value. Publications that contributed software tools or techniques to encourage people to be helpful towards each other were classified by the raters as relevant to \textit{Helpful}. 

\vspace{0.25cm}
Moreover, the second highest number of relevant publications was observed for \textit{Privacy} (Figure~\ref{fig:specific-values-occurrences}). This category contained papers that directly considered user privacy. Also, \textit{Protecting the environment}, the third most commonly found value, appeared in publications that directly considered \textit{Sustainability} and \textit{Energy efficiency} in software. 

\vspace{0.5cm}

\begin{figure}
    \centering
    \includegraphics[width=0.5\textwidth]{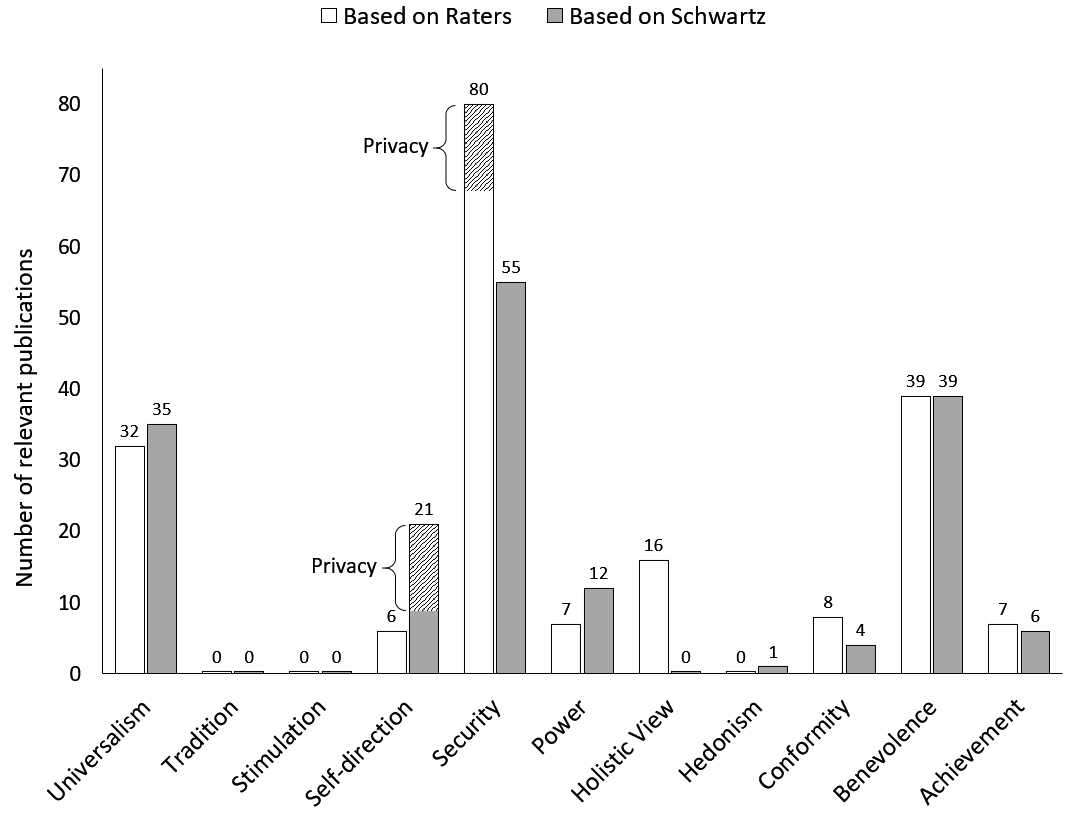}
    \caption{Considering value categories in SE publications}    
    \label{fig:value-categories-side}
\end{figure}

\textit{Which value categories are commonly considered?} As explained in section \ref{subsec:classification-process}, the raters were given the freedom to classify the publications under different value categories regardless of SVS hierarchical structure. As a result, the raters were allowed to pick, for a publication, values and value categories that did not necessarily match in SVS. Figure~\ref{fig:value-categories-side} shows the prevalence of the publications under different value categories specified by the raters. 

Figure~\ref{fig:value-categories-side} also shows how those papers would have been classified if the raters strictly followed SVS (Figure~\ref{fig:values}): a significant difference was observed for value categories \textit{Security} and \textit{Self direction}, where the raters classified 80 papers as relevant to \textit{Security}; had the raters followed SVS for classification, only 55 papers would have been classified under \textit{Security}. On the other hand, raters classified only 6 papers as relevant to \textit{Self direction}. If it was based on SVS, 21 papers would have fallen under the category of \textit{Self direction} (Figure~\ref{fig:values}). 

Scrutinizing the publications classified under \textit{Security} and \textit{Self direction} revealed an interesting finding: the raters chose \textit{Security} as the category of 12 papers classified as relevant to \textit{Privacy}, but based on Schwartz Values Structure (SVS), \textit{Privacy} is under \textit{Self direction}. As such, those 12 papers (annotated on the graph of Figure~\ref{fig:value-categories-side}) could have been classified under \textit{Self direction} if SVS (Figure~\ref{fig:values}). Though relatively small, similar differences were also observed for other value categories such as \textit{Power}, \textit{Achievement}, \textit{Conformity}, and \textit{Hedonism}. Considering the SE background of most of the raters (Section~\ref{sec:methodology}), this raised a major question: ``do software engineers perceive values differently from social scientists?'' To reflect the view of the raters in our discussions, we use, consistently, the categories specified by them in the rest of the paper.

\begin{figure}[htb]
    \centering
    \includegraphics[width=0.45\textwidth]{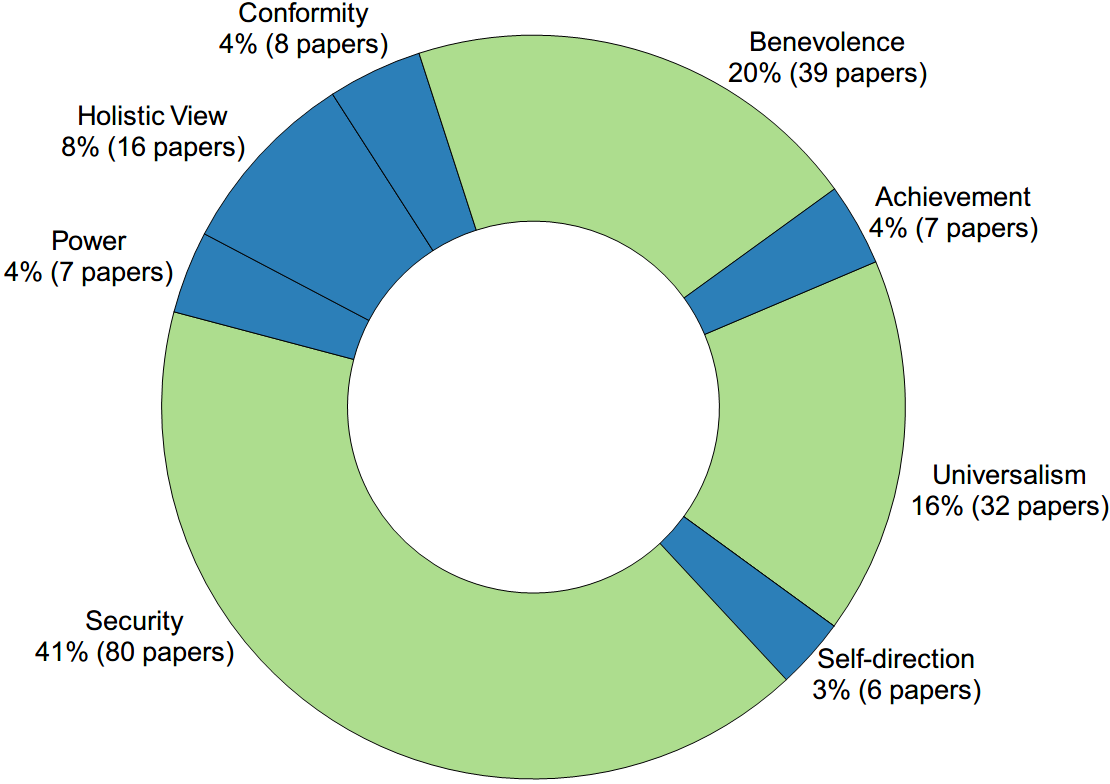}
    \caption{Relevant publications per value category}
    \label{fig:value-categories-proportion}
\end{figure}

It can be observed from Figure~\ref{fig:value-categories-proportion} that 80 papers (41\% of the relevant publications) were classified as relevant to \textit{Security}, which made \textit{Security} the most prevalent value category. This was not hard to predict as \textit{Security} is a well-recognized quality aspect of software, for which there is a great demand from stakeholders. The second and third most highly prevalent value categories were found to be \emph{Benevolence} and \emph{Universalism}, which constituted 20\% and 16\% of the relevant publications, respectively. On the other hand, no publications were found to be relevant to the categories \emph{Tradition}, \emph{Stimulation}, and \emph{Hedonism}. Moreover, 8\% of the relevant papers were classified under the category \textit{Holistic view}, which does not exist in SVS -- this category was introduced based on the raters' feedback from the pilot study (Section~\ref{subsec:Pilot-Phase}) to account for publications that considered values in general.  
\subsection{Relevant Publications per Venue}
\label{subsec:values-distribution}

\begin{figure}[htb]
    \centering
    \includegraphics[width=0.45\textwidth]{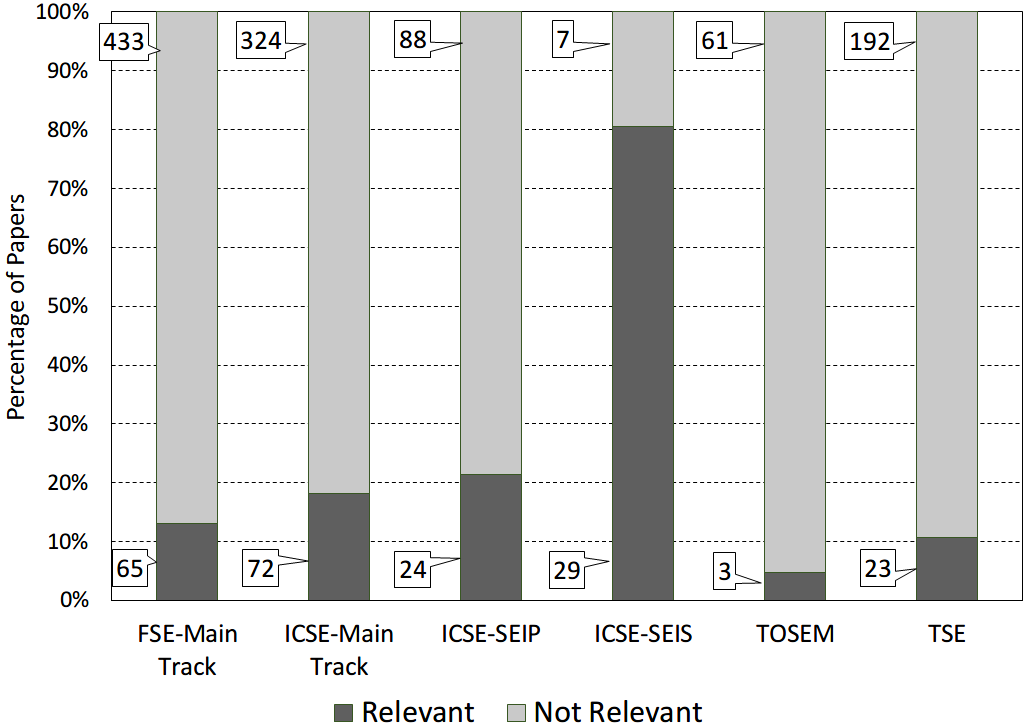}
    \caption{Proportion of values relevant publications in SE venues/tracks. The labels on the bars denote the number of papers in each category. }
    \label{fig:venue-relevance}
\end{figure}

To answer \textbf{(RQ3)}, this section reports our findings on the distribution of values relevant to SE publications across SE venues. Figure~\ref{fig:venue-relevance} demonstrates, for each venue/track, the proportion of the relevant publications in 2015 -- 2018. 

\textit{The proportion of relevant publications in each venue/track}. We observed (Figure~\ref{fig:venue-relevance}) that the proportion of  relevant publications in the SE journals, namely TOSEM (about 5\%) and TSE (about 11\%), is lower than the proportion of relevant publications in the main tracks of ICSE (about 18\%) and FSE (about 13\%), and significantly lower than the proportion of relevant papers in the SEIP (21\%) and SEIS (about 81\%) tracks of ICSE. In particular, the proportion of values relevant papers was significantly higher in SEIS. This is not surprising given the focus of the track. 

\begin{figure}[htbp]
    \centering
    \includegraphics[width=0.45\textwidth]{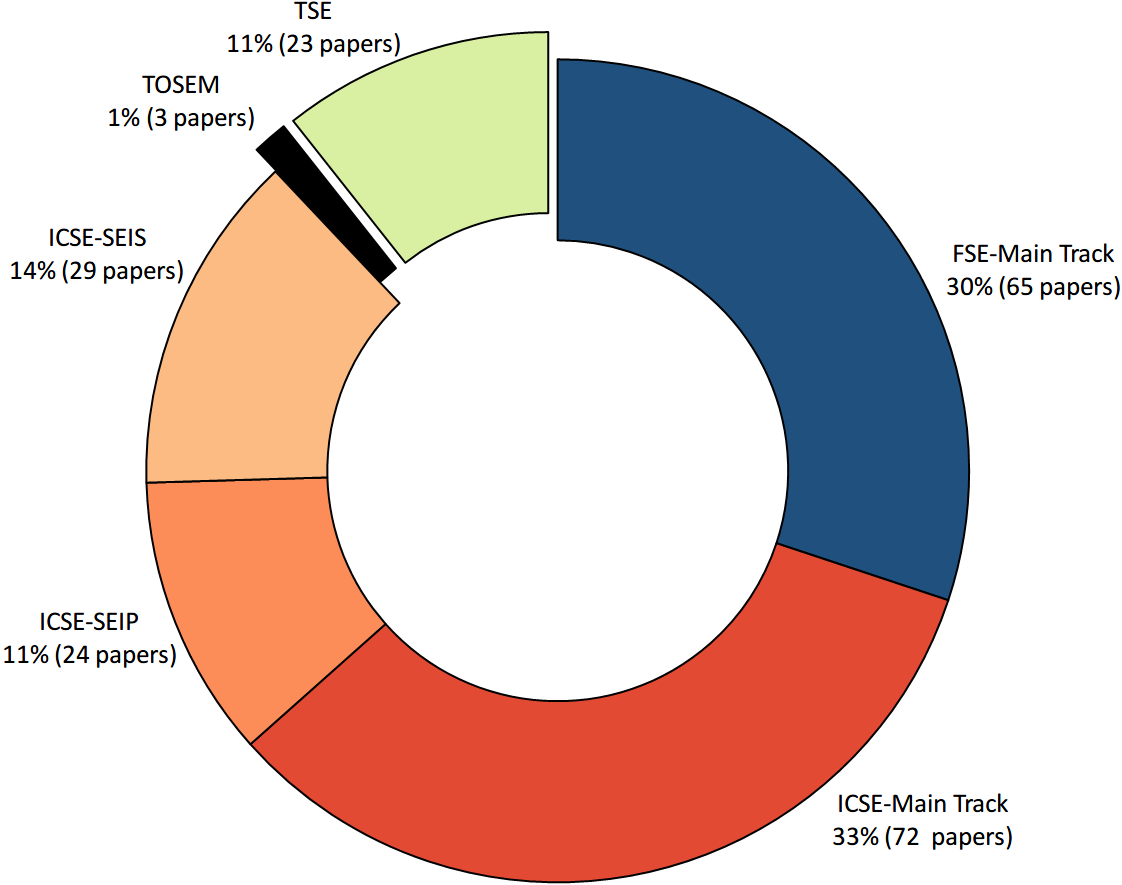}
    \caption{Relevant publications per venue/track}
    \label{fig:venue-relevance-distribution}
\end{figure}

\textit{The distribution of  relevant publications by venue/track}. Figure \ref{fig:venue-relevance-distribution} demonstrates the distribution of relevant publications across the studied venues/tracks. From all 216 publications that directly considered values (relevant publications), 58\% were published in different tracks of ICSE: main track (33\%), SEIS (14\%), and SEIP (11\%). The highest prevalence of relevant publications was seen in the main tracks of ICSE (33\%) and FSE (30\%). As such, it was concluded that about $88\%$ of the publications that directly considered values were published in SE conferences: ICSE (58\%) and FSE (30\%). On the other hand, SE journals, TSE (11\%) and TOSEM (1\%),  constituted only 12\% of the relevant publications (Figure~\ref{fig:venue-relevance-distribution}). 

\begin{figure*}[htbp]
    \centering
    \includegraphics[width=\linewidth]{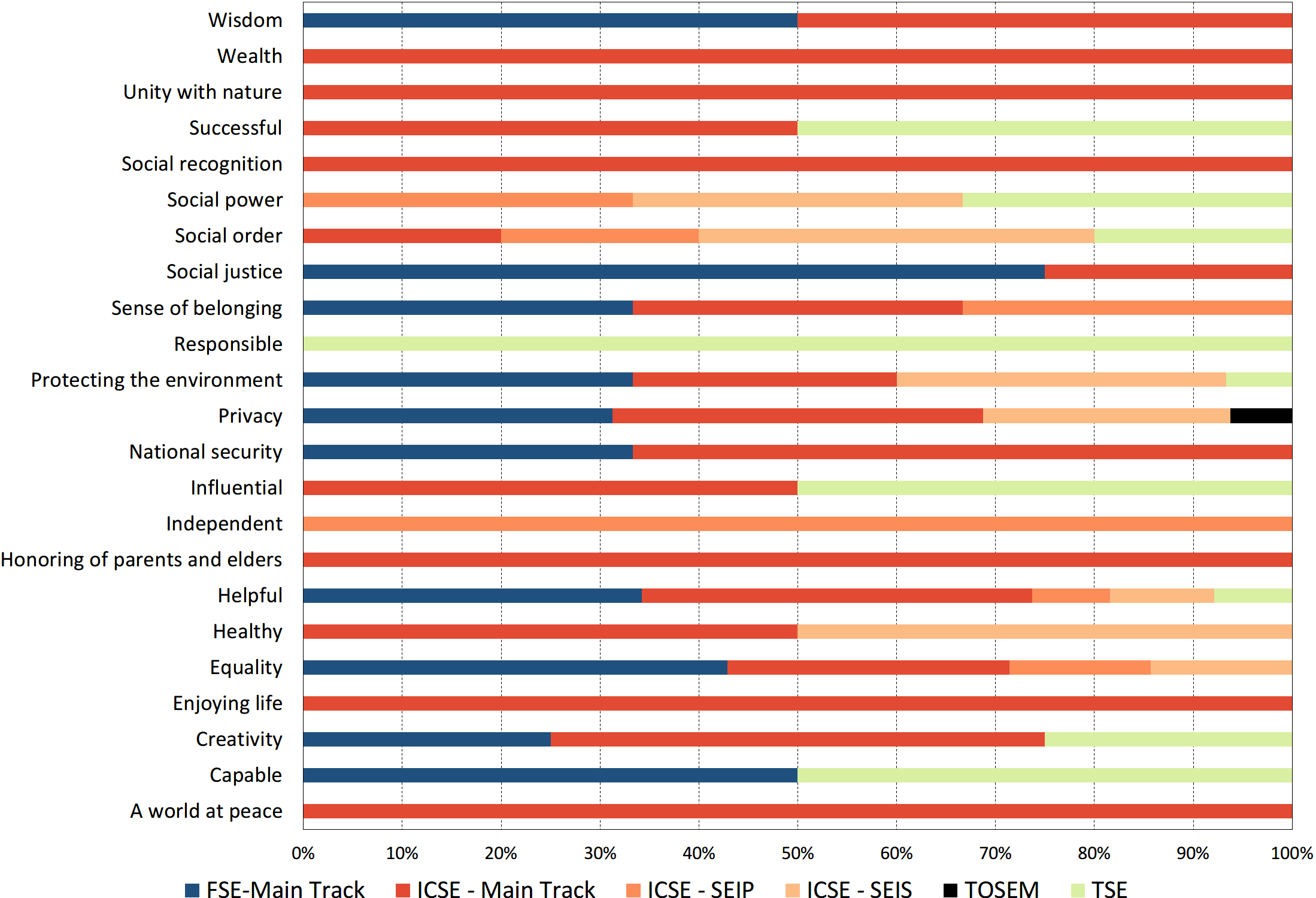}
    \captionsetup{margin=16ex}
    \caption{The distribution of publications relevant to different values by venue/track; relevant publications were found only for 23 out of 58 values in Schwartz Value Structure (Figure~\ref{fig:values}).}
    \label{fig:individual-by-venue}
\end{figure*}

\textit{The distribution of relevant publications by values and venues}. Figure~\ref{fig:individual-by-venue} shows how the publications relevant to different values are distributed across different venues/tracks. We observed that only 23 out of 58 values in SVS (Figure~\ref{fig:values}) were considered. For some values, relevant publications were found across most venues/tracks; publications relevant to \textit{Helpful} were found in 5 out of 6 venues/tracks. But for the majority of the considered values in Figure~\ref{fig:individual-by-venue} (15 out of 23), the number of the venues/tracks that published papers relevant to those values did not exceed 2. For instance, publications relevant to \emph{Social justice} and \emph{National security} were found only in the main tracks of FSE and ICSE. Also, publications relevant to \emph{Enjoying life}, \emph{Honoring of the parents and elders}, and \emph{A world at peace} appeared only in the main track of ICSE. Also, publications relevant to certain values, e.g. \emph{Equality}, \emph{Social justice}, and \textit{Healthy}, were only present in conference papers but not in journals. We further observed that for the majority of  values (19 of 23 values in Figure~\ref{fig:individual-by-venue}), relevant publications were found in the main track of ICSE while publications in TOSEM only considered \textit{Privacy}.

\begin{figure}[htb]
    \includegraphics[width=0.5\textwidth]{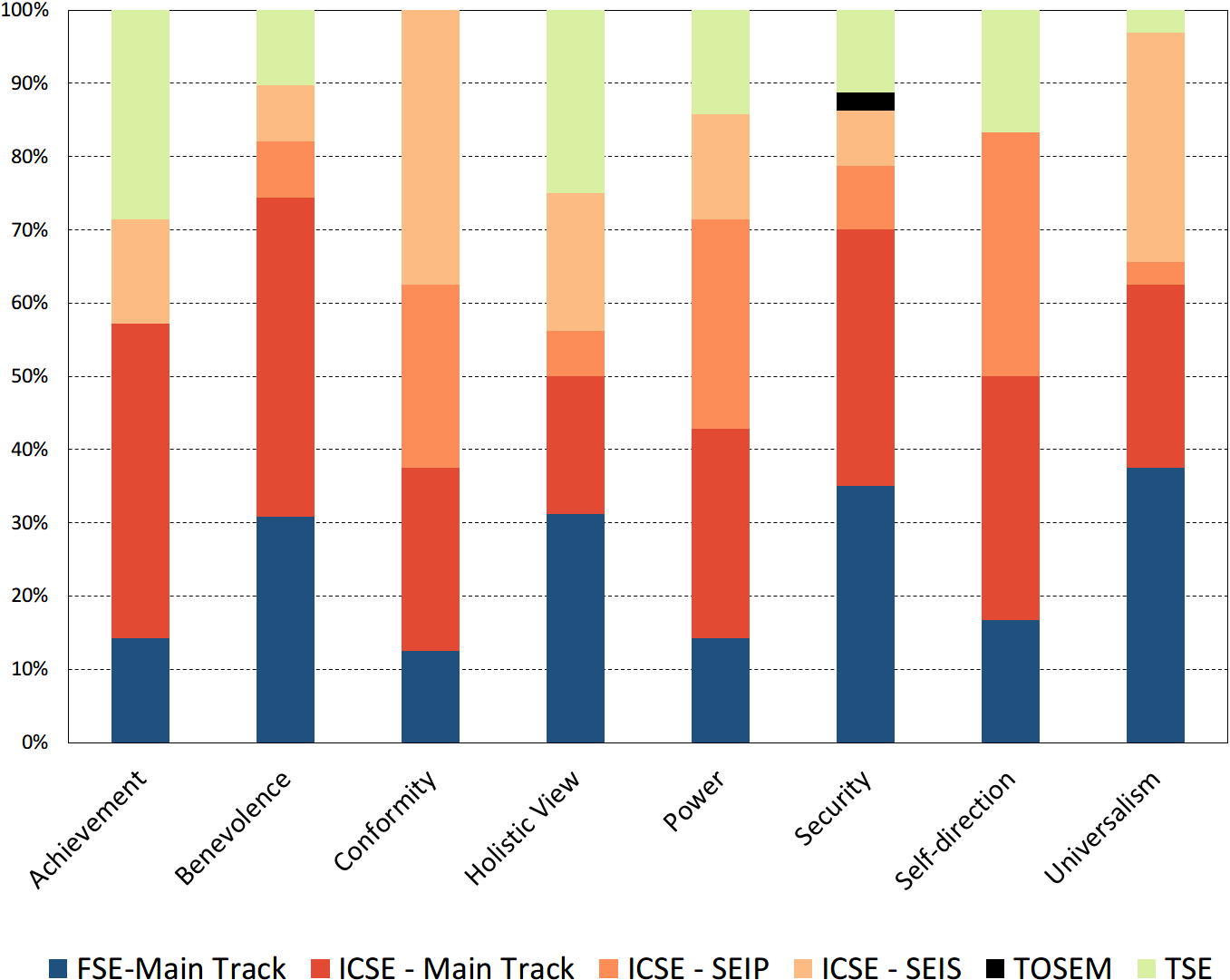}
    \captionsetup{margin=4ex}
    \caption{Publications relevant to different value categories across SE venues/tracks.\davoud{is it possible. if it's not gonna impact the gray scale printed version, to choose a different color for TOSEM? e.g. black; currently it looks like a gap on the bar as it is the same color as the background}}
    \label{fig:venue-value-category}
\end{figure}

\textit{The distribution of relevant publications by value categories and venues}. Publications relevant to  7 out of 10 value categories in SVS (Figure~\ref{fig:values}) were found across different venues/tracks (in Figure~\ref{fig:venue-value-category}). We further found publications relevant to category \textit{Holistic view}, which was introduced based on pilot study, as discussed in Section~\ref{subsec:Pilot-Phase}. Publications relevant to all these 8 value categories were found in the main tracks of FSE and ICSE (Figure~\ref{fig:venue-value-category}). Also, publications relevant to \emph{Security} were found in all SE venues. Moreover, publications that directly considered \emph{Benevolence} and \emph{Universalism} were found across most venues/tracks. Publications relevant to \emph{Universalism} were more prevalent in the SEIS track of ICSE. However, publications in TOSEM only considered \emph{Security} but not other value categories. It was also interesting to see that, compared to other venues/tracks, the SEIS track of ICSE contained the highest proportion of publications relevant to \textit{Conformity}.

\subsection{Data Availability}
The dataset that supports the findings of this study is available at https://figshare.com/s/7a8c55799584d8783cd6.

\section{Discussion}

We carried out this research to verify our hypothesis that SE research does not sufficiently consider human values. Our research findings confirm this intuition. The extent to which SE research ignores human values is significant: 1105 out the selected 1350 papers (82\%) were found not to be relevant to human values. Furthermore, out of 195 papers that do address values, 80 papers relate to \textit{Security}. This is unsurprising, but also illustrates that the lack of consideration of human values in SE is even more stark. Indeed, a majority of other human values (approximately 79\%) are not adequately addressed in SE research. 

The value of \textit{Helpful} that relates to ``preserving and enhancing the welfare of those with whom one is in frequent personal contact'' (Table~\ref{tab:valuecategories_defOnly}) was the highest classified among all (58) values. This suggests that SE research is often aimed at being helpful to the SE community -- for example, by means of improving processes thus reducing development effort or removing development obstacles, or developing new tools and techniques to facilitate or improve certain practices or tasks. Our results indicate that only a small proportion of publications  relate to individualistic-value categories (\textit{Hedonism, Achievement, Stimulation and Power}) compared to group-value categories (\textit{Universalism, Conformity and Tradition}). This verifies the tensions discussed in the Schwartz Values Structure about the competing and contradicting nature of these bipolar value categories~\cite{schwartz1994there}. 
  
It is important to note that SVS served as an appropriate yet not ideal scheme for classifying human values in SE. We discovered that SVS does not include some values commonly discussed in SE. For example, sustainability is a  value that has received significant recent attention in SE, yet is not listed among the 58 SVS values. Since SVS originates in the social sciences, raters sometimes found it difficult to map certain SE values to SVS-prescribed value categories. This is likely due to the difference in meaning of values in different contexts (i.e., social sciences versus software engineering). Future work will look at how to adapt SVS to an SE context.


Without attempting to generalize, certain findings are worth mentioning here. For example, among the selected venues, ICSE has the most diverse range of values covered compared to others. In addition, there are certain values such as \textit{Wealth}, \textit{Unity with nature}, \textit{Social recognition}, \textit{Honoring of parents and elders}, \textit{Enjoying life}, and \textit{A world at peace} found in ICSE publications but not addressed in any other venue. It is difficult to attribute this to a trend in ICSE submissions or to the broad nature of ICSE. Similarly, for other venues, a broader and more comprehensive  study is needed to discuss any trends. 

\section{Threats to Validity} 

In this section we discuss limitations of this research categorized as \textit{Internal}, \textit{External} and \textit{Construct} validity threats.

\textit{{\textbf{Construct Validity}:} }Choosing a classification scheme suited for the software engineering domain was one of the main challenges for this research. In the absence of an SE-specific scheme to classify human values, we selected the Schwartz Values Structure (SVS). SVS is a well established theory for understanding human values in the social sciences. It has been successfully applied in Human and Computer Interaction (HCI) and Information and Computer Technologies (ICT) to study and explain human values~\cite{thew2018value}. Using SVS as an independent classification scheme, instead of developing our own, mitigated the risk of introducing researcher bias. 

Similar to Glass et al. \cite{glass2002research}, lack of mutual exclusion was a challenge for our classification scheme. It was often possible to classify a paper as relating to more than one individual value. This we believe was more to do with the ill-defined nature of human values  than a limitation of the chosen classification scheme. Still, the potential threat was mitigated by using an iterative process and conducting rater training to understand and clarify relationships between values and their categories.

In some cases, the raters found that certain papers related to human values in general rather than any particular value. Forcing such papers into a single value category would have influenced results. To mitigate this, we added a new \emph{Holistic view} category. Some papers relating to \textit{Privacy} were categorized under \emph{Security} rather than \emph{Self direction}. Raters, based on their understanding of \emph{Privacy} in an SE context, considered it to be more relevant to \emph{Security} than \emph{Self direction}. This may have influenced the results. To mitigate this, we provided results for both rater preferred and SVS prescribed categories in Figure~\ref{fig:value-categories-side}. Some common SE values were not found in SVS, such as \emph{Sustainability}. 

SVS may not be the ideal classification scheme for SE, and we expect useful further research to adapt SVS to SE context.

\textit{\textbf{Internal Validity}} threats for this study arise from the complexities of categorizing papers into the selected classification scheme. It is possible that the raters' own expertise in understanding the scheme categories and definitions of values may have influenced paper classifications. This risk however, was mitigated as the classification process forced random assignment of each paper to two raters and in case of a disagreement an independent arbiter was introduced to facilitate agreement. 
Some disagreements (2\%, see Figure~\ref{fig:relevant})  remained even after the arbiter's intervention.\waqar{(@Arif can you provide percentage here, and also the figure Fig abc)} \arif{I put the one on the Directly Relevant level. Do you also need for the category and value also?}\waqar{need to discuss this in person} In such cases we did not force consensus.

\textbf{\textit{External validity} }threats may arise from potential limitations  of our choice of publication venues and the block of time period under study (i.e., 2015-2018). The chosen venues are widely acknowledged as the top-tier venues of SE research; however, we accept that the results may be different if other more specialist conferences/journals had been considered.

Generalizability of results based on a subset of papers is often a concern for empirical studies.  In our research, this risk was mitigated by using 1350 papers published in the last 4 years which can be considered a good representation of trends in SE research as suggested in \cite{Bertolino2018}. The findings of this study, however, may be biased towards ICSE and FSE  as they published more papers in the selected period compared to journals (ICSE 559, FSE 512 vs. TSE 215 and TOSEM 64).

While a detailed review of the entire papers (rather than just the abstract, title and keywords) could have provided more accurate results, we adopted a procedure similar to those used in previous studies \cite{shaw2003writing,Bertolino2018}. The time required for reliable classification means that reading all 1350 papers is infeasible.





\section{Related Work}
\label{sec:related-work}
Classification of papers has been widely adopted in the SE literature~\cite{shaw2003writing,systa2012inbreeding,montesi2008software,vessey2002research} as a way of providing insights on trends and directions in SE research. Such findings, though not conclusive, can indicate the general attitude of SE researchers as well as the priorities in SE research. Paper classification helps to highlight the gaps and the needs for further research in specific SE domains. Mary Shaw~\cite{shaw2003writing}, for instance, analyzed the abstracts of research papers submitted and accepted to ICSE 2002 to identify different research types as well as the trends in research question types, contribution types and validation approaches. The author also studied the program committee discussions regarding the acceptance or rejection of the papers. Another example is the work by Vessey et al.~\cite{vessey2002research}: to report their findings, the authors categorized samples of SE papers published from 1995 to 1999 in six journals based on topic, method, and approach. 

However, paper classification methods rely on classification schemes, that can be general or specific depending on the purpose of the classification. To classify different SE papers, Montesi and Lago~\cite{montesi2008software} presented a paper classification approach based on the call for papers of top-tier SE conferences and journals included in the Journal Citation Reports and the instructions to authors of relevant journals and published works. Also, Ioannidis et al.~\cite{ioannidis2015meta} categorized the meta-research discipline into five main thematic fields corresponding to how to conduct, report, verify, correct and reward science. There have also been efforts to develop specific classification schemes. For instance, Wieringa et al.~\cite{wieringa2006requirements} developed a classification scheme to identify papers that belong to Requirements Engineering as a subdomain in SE. Sjoberg et al.~\cite{sjoberg2005survey} surveyed SE papers in nine journals and three conferences from 1993 to 2002 with the aim to characterize controlled experiments in SE by characterizing the topics of the experiments and their subjects, tasks, and environments. 

Moreover, some paper classifications have identified gaps in SE practice. An example is the work by Stol and Fitzgerald~\cite{stol2015holistic}, where the authors observed the lack of a holistic view in SE research. The work contributed a framework for positioning a holistic set of research strategies and showed its strengths and weaknesses in relation to various research components. Also, Zelkowitz and Wallace~\cite{zelkowitz1997experimental} classified, according to a 12-model classification scheme, around 600 SE papers published over a period of three years to provide insights on the use of experimentation within SE. They identified a gap in SE research with respect to validation and experimentation. Another example is an empirical study of SE papers performed by Zannier et al.~\cite{zannier2006success} to investigate the improvement of the quantity and quality of empirical evaluations conducted within ICSE papers over time. The authors compared a random sample of papers in two periods, 1975 -- 1990 and 1991 -- 2005, and found that the quantity of empirical evaluation has grown, but the soundness of evaluation has not grown at the same pace.

Last but not least, some paper classifications have provided insights on SE venues in relation to the papers published in those venues. An example is the work by Systa et al.~\cite{systa2012inbreeding} that investigated the turnover of PC compositions and paper publication in six SE conferences. The work was later extended by Vasilescu et al. ~\cite{vasilescu2014healthy} by proposing a wider collection of metrics to assess the health of 11 SE conferences over a period of more than 10 years. 
\section{Conclusions and Future Work}
\label{sec:conclusion}


Repeated incidents of software security and privacy violations continue to attract researchers' attention. In this paper, however, we investigated the prevalence of a broader range of human values including \textit{Trust}, \textit{Equality} and \textit{Social justice} in software engineering research. Using Schwartz Values Structure as our classification scheme, we classified 1350 recently published (2015--2018) papers in top-tier SE conferences and journals. 
We conclude that only a small proportion of SE research considers human values. While \textit{Security}, as a value category, and \textit{Privacy}, as a specific value, stand out as the main focus in SE research, few other human values such as \textit{Helpful}, \textit{Protecting the environment} and \textit{Social justice} are considered. A broad range of human values remain inadequately addressed in SE research. Finally, we found SE conferences publish more values relevant research compared to SE journals. 

In future, we would like to extend this study using a machine learning approach. Manually labelled data from this study could be used for training machine learning algorithms to classify larger sets of publications with the aim to better visualize how SE research addresses human values. We also plan to utilise our manually labelled data captured from various SE contexts to develop definitions of human values that are relatively easy for practitioners to understand and implement. Finally, we plan to carry out case studies in software organizations to investigate whether SE research related to human values has actually made an impact on SE practice.


\bibliographystyle{ACM-Reference-Format}
\bibliography{ms}
\end{document}